 \documentclass[11pt]{article}

\usepackage{amsmath,amsfonts,amssymb}

 \usepackage[numbers,sort&compress]{natbib}

\usepackage{doi}
\renewcommand{\doitext}{\noexpand\textsc{doi:}}

\usepackage{hyperref}
\hypersetup{
   colorlinks   =  true,
    linkcolor    = blue,
    citecolor    = red,
     urlcolor	=blue,     
	urlbordercolor = {1 1 0}
}

\newcommand\be{\begin{equation}}
\newcommand\ee{\end{equation}}
\newcommand\bea{\begin{eqnarray}}
\newcommand\eea{\end{eqnarray}}

\usepackage{lmodern}
\usepackage[english]{babel}
\usepackage[utf8]{inputenc}
\usepackage[T1]{fontenc}

\newcommand\RR{\mathbb{R}}
\newcommand\CC{\mathbb{C}}

\newcommand{\rd}{\mathrm{d}} 
\newcommand{\re}{\mathrm{e}} 
\newcommand{\T}{\mathsf{T}}

\newcommand{\pdv}[1]{\frac{\partial}{\partial #1}}
\newcommand\semicolon{\nobreak\mskip2mu\mathpunct{}\nonscript\mkern-\thinmuskip{;}%
\mskip6muplus1mu\relax}
\DeclareMathOperator\sign{sign}

\begin{document}
\
  \vskip0.5cm

  \begin{center}     
 \noindent
 {\Large \bf  
  Lie--Hamilton systems associated with the\\[6pt] symplectic Lie algebra $\mathfrak{sp}(6, \RR)$} 
 
   \end{center}

\medskip

\begin{center}

{\sc  Oscar Carballal$^{\dag}$\footnote{Based on the contribution presented at the
    ``XXIVth International Conference on Geometry, Integrability and Quantization'' held in Varna, Bulgaria, June 6--13, 2024}, Rutwig Campoamor-Stursberg$^{\dag,\ast}$  and Francisco J.~Herranz$^\ddag$}

\end{center}

\medskip

\noindent 
$^\dag$ Departamento de \'Algebra, Geometr\'{\i}a y Topolog\'{\i}a,  Facultad de Ciencias 
Matem\'aticas,\\ Universidad Complutense de Madrid, Plaza de Ciencias 3, E-28040 Madrid, Spain

  \noindent
$^\ast$ Instituto de Matem\'atica Interdisciplinar, Universidad Complutense de Madrid,\\ E-28040 Madrid,  Spain

\noindent
{$^\ddag$ Departamento de F\'isica, Universidad de Burgos, 
E-09001 Burgos, Spain}

 \medskip
 
\noindent  E-mail: {\small
\href{mailto:oscarbal@ucm.es}{\texttt{oscarbal@ucm.es}},  \href{mailto:rutwig@ucm.es}{\texttt{rutwig@ucm.es}},  \href{mailto:fjherranz@ubu.es}{\texttt{fjherranz@ubu.es}} 
}
 
\medskip

\begin{abstract}
\noindent
New classes of Lie--Hamilton systems are obtained from the six-dimensional fundamental representation of the symplectic Lie algebra $\mathfrak{sp}(6,\mathbb{R})$. The ansatz is based on a recently proposed procedure for constructing higher-dimensional Lie--Hamilton systems through the representation theory of Lie algebras. As applications of the procedure, we study a time-dependent electromagnetic field and several types of coupled oscillators. The irreducible embedding of the special unitary Lie algebra $\mathfrak{su}(3)$ into $\mathfrak{sp}(6,\mathbb{R})$ is also considered, yielding Lie--Hamilton systems arising from the sum of the quark and antiquark three-dimensional representations of $\mathfrak{su}(3)$, which are applied in the construction of $t$-dependent coupled systems.  In  addition, $t$-independent constants of the motion are obtained explicitly for all these Lie--Hamilton systems, which allows   the derivation of a nonlinear superposition rule
\end{abstract}
\medskip
\medskip

  \noindent
\textbf{Keywords}: constants of the motion,   coupled   oscillators, Lie systems, Minkowski, nonlinear differential equations,  nonlinear superposition rules, special unitary Lie algebra, symplectic Lie algebra    
 
   \medskip

\noindent
\textbf{MSC 2020 codes}: 17B10, 34A26, 34C14, 58A30

 \newpage

 \tableofcontents


\section{Introduction}

Group theory has been shown to be a very effective method within the theory of differential equations, helping to identify symmetry schemes and linearization properties, as well as effective criteria to derive (nonlinear) superposition rules \cite{Blu,Olv,Win}. The combination of Lie groups with additional geometric structures, on the other hand, has enlarged the number of techniques to analyze the existence and computation of superposition rules for differential equations, hence enlarging the original approach of S. Lie \cite{Lie}. In this context, the theory of Lie--Hamilton systems (LH in short) has been shown to be an effective approach, with wide applications to both classical and quantum systems (see e.g. \cite{deLucas2020} and references therein). In this context, genuinely indecomposable LH systems in higher dimension, i.e., systems that cannot be decoupled into systems in lower dimension, have recently been introduced and analyzed in connection with the representation theory of (semisimple) Lie algebras \cite{Campoamor2024}, leading to the construction of new intrinsic hierarchies of higher-dimensional LH systems. In this paper, we generalize the construction done in \cite{Campoamor2024} for the symplectic Lie algebra $\mathfrak{sp}(4,\mathbb{R})$ to the rank-three case, i.e., the Lie algebra $\mathfrak{sp}(6,\mathbb{R})$, hence deriving new three-dimensional LH systems related to coupled oscillators.

This paper is structured as follows. After presenting some generalities on Lie and LH systems in Section~\ref{Generalities}, we extend  in Section~\ref{sec:1} the formalism developed  in \cite{Campoamor2024} based on the Lie algebra $\mathfrak{sp}(4, \RR)$ to the case of the Lie algebra $\mathfrak{sp}(6, \RR)$, leading to novel LH systems on $\T^{*}\RR^{3}$. Furthermore, in Subsection~\ref{subsection:constants} their corresponding $t$-independent constants of the motion are obtained in an explicit form following the  coalgebra formalism introduced in~\cite{constants2021,constants2013}. These can then be used to deduce a   nonlinear superposition rule which is also indicated.   As applications of such $\mathfrak{sp}(6, \RR)$-LH systems, we construct a $t$-dependent electromagnetic field and several coupled oscillators in Subsection~\ref{subsection:applicationssp6}, which generalize the systems obtained in \cite{Campoamor2024}.  In Section~\ref{section:su3} we study LH systems based on the regular subalgebra $\mathfrak{su}(3) \subset \mathfrak{sp}(6, \RR)$, which allow the construction of  coupled systems  presented in Subsection~\ref{subsection:appsu3}. In particular, we obtain  an $\mathfrak{su}(3)$-LH system that can be interpreted as a coupling of two LH systems,  each of them being a coupled oscillator on the Minkowski plane $\T^{*}\mathbf{M}^{1+1}$.
 The paper finishes with some concluding remarks and future work prospectives.


\section{Basics on Lie and Lie--Hamilton systems}
\label{Generalities}
The solutions of a non-autonomous first-order system of ordinary differential equations (ODEs) on a smooth manifold $M$ in normal form 
\begin{equation}
\frac{\rd x^{j}}{\rd t} = \psi^{j}(t, \mathbf{x}), \qquad \mathbf{x} \in M, \qquad 1 \leq j \leq n = \dim M
\label{eq:intro:system}
\end{equation}
are given by the integral curves of a $t$-dependent vector field $\mathbf{X}: \RR \times M \to \T M$, the local expression of which is given by 
\begin{equation}
\mathbf{X} (t, x) := \psi^{j}(t, \mathbf{x}) \pdv{x^{j}}.
\nonumber
\end{equation}
This allows us to univocally identify the system \eqref{eq:intro:system} with the $t$-dependent vector field $\mathbf{X}$. We say that \eqref{eq:intro:system} is a \textit{Lie system} if its general solution $\mathbf{x}(t)$ can be written in terms of a $t$-independent function $\Psi$, a \textit{superposition rule}, as
\begin{equation}
\mathbf{x}(t) = \Psi(\mathbf{x}_{1}(1), \ldots \mathbf{x}_{r}(t); k_{1}, \ldots, k_{n})
\nonumber
\end{equation}
where $\mathbf{x}_{1}(t), \ldots, \mathbf{x}_{r}(t)$ denotes a generic finite family of particular solutions and $(k_{1}, \ldots, k_{n}) \in M$ a certain point related to the initial conditions (for general aspects of the theory of Lie systems, see e.g. \cite{Car,deLucas2020} and references therein).

  The fundamental Lie--Scheffers Theorem states that a system \eqref{eq:intro:system} is a Lie system if and only if the $t$-dependent vector field $\mathbf{X}$ can be decomposed as 
\begin{equation}
\mathbf{X}(t, \mathbf{x}) = \sum_{i = 1}^{\ell} b_{i}(t)\mathbf{X}_{i}(x) 
\nonumber
\end{equation}
for some functions $b_{1}(t), \ldots, b_{\ell}(t)$ and vector fields $\mathbf{X}_{1}, \ldots, \mathbf{X}_{\ell}$ on $M$ spanning an $\ell$-dimensional real Lie algebra $V^{X}$ of vector fields, the so-called \textit{Vessiot--Guldberg (VG) Lie algebra} \cite{Lie}. This allows to regard a Lie system as a curve taking values values in a finite-dimensional Lie algebra $V^{X}$ of vector fields. 

One of the main features of Lie systems is that they can be made compatible with a lot of supplementary geometric structures (Jacobi, Poisson, symplectic, multi-symplectic, Dirac, contact, etc.), a fact that provides additional properties of notable interest for both the quantitative and qualitative analysis of the associated systems of ODEs and, in some cases, facilitates the deduction of an explicit superposition rule. 

For the particular case of symplectic geometry, a Lie system $\mathbf{X}$ on a symplectic manifold $(M, \omega)$ is called a \textit{LH system} if it admits a VG Lie algebra $V^{X}$ formed by Hamiltonian vector fields with respect to the symplectic form $\omega$. Moreover, if $\mathbf{X}_{1}, \ldots, \mathbf{X}_{\ell}$ span  $V^{X}$, their associated Hamiltonian functions $h_{1}, \ldots, h_{\ell} \in C^{\infty}(M)$, determined by the inner product condition 
\begin{equation}
\iota_{\mathbf{X}_{i}} \omega = \rd h_{i}, \qquad 1 \leq i \leq \ell ,
\label{inner}
\end{equation} 
span a finite-dimensional (functional) real Lie algebra $\mathcal{H}_{\omega}$ with respect to the Poisson bracket $\{\cdot, \cdot\}_{\omega}$ induced by the symplectic form $\omega$, known as a \textit{LH algebra} of $\mathbf{X}$. At this point, we recall that the symmetric algebra $S(\mathcal{H}_{\omega})$ of $\mathcal{H}_{\omega}$ can be endowed with a Poisson--Hopf algebra structure which allows an algorithmic computation of $t$-independent constants of the motion, hence eventually simplifying the deduction of a superposition rule~\cite{constants2021,constants2013}.


\section{The symplectic Lie algebra $\mathfrak{sp}(6, \RR)$}
\label{sec:1}
The real symplectic Lie algebra $\mathfrak{sp}(6, \RR)$ is defined by the matrix constraint 
\begin{equation}
\mathfrak{sp}(6, \RR) := \{X \in \mathfrak{gl}(6, \RR)\semicolon  X^{T}J + JX = 0\}
\nonumber
\end{equation}
where $J$ is the matrix $J := \begin{pmatrix}
 0 & \mathrm{Id}_{3} \\
 - \mathrm{Id}_{3} & 0
\end{pmatrix}$. Instead of using the usual basis of $\mathfrak{sp}(6, \RR)$ described through elementary matrices, we consider the so-called boson basis \cite{Lip,Hav}, as done in \cite{Campoamor2024} for the construction dealing with the Lie algebra $\mathfrak{sp}(4, \RR)$. 

In terms of the creation $a_{i}$ and annihilation $a_{i}^{\dagger}$ operators $(1 \leq i \leq 3)$, the  Lie algebra $\mathfrak{sp}(6, \RR)$ is spanned by the operators $a_{i}^{\dagger}a_{j}$, $a_{i}^{\dagger}a_{j}^{\dagger}$ and $a_{i}a_{j}$. We label this boson basis as 
\begin{equation}
X_{i,j} := a_{i}^{\dagger} a_{j}, \qquad X_{-i,j} := a_{i}^{\dagger} a_{j}^{\dagger}, \qquad X_{i,-j}:= a_{i}a_{j}
\nonumber
\end{equation}
in such a way that the constraint 
\begin{equation}
X_{i,j} + \varepsilon_{i} \varepsilon_{j} X_{-j,-i}= 0, \quad \varepsilon_{i} := \sign(i), \quad \varepsilon_{j}:= \sign(j), \quad -3 \leq i, j \leq 3
\nonumber
\end{equation}
is satisfied. The commutation relations of $\mathfrak{sp}(6, \RR)$ over this basis are 
\begin{equation}
[X_{i,j}, X_{k, \ell}] = \delta_{j,k} X_{i ,\ell} - \delta_{i ,\ell} X_{k,j} + \varepsilon_{i} \varepsilon_{j} \delta_{j,-\ell} X_{k,-i}- \varepsilon_{i}\varepsilon_{j} \delta_{i,-k} X_{-j,\ell}
\label{eq:sp6:comm}
\nonumber
\end{equation}
where $-3 \leq i, j, k, \ell \leq 3$.

Let us   consider the six-dimensional fundamental (defining) representation $\Gamma_{\omega_{1}}: \mathfrak{sp}(6, \RR) \to \mathfrak{gl}(6, \RR)$ of $\mathfrak{sp}(6,\RR)$ given in matrix form by 
\begin{equation}
A_{\Gamma_{\omega_{1}}} = \begin{pmatrix}
X_{1,1} & X_{1,2} & X_{1,3} & - X_{-1,1} & - X_{-1,2} & -X_{-1,3} \\
X_{2,1} & X_{2,2} & X_{2,3} & -X_{-1,2} & - X_{-2,2} & - X_{-2,3} \\
X_{3,1} & X_{3,2} & X_{3,3}& - X_{-1,3} & - X_{-2,3} & - X_{-3,3} \\
X_{1,-1} & X_{1,-2} & X_{1,-3} & - X_{1,1} & -X_{2,1} & - X_{3,1} \\
X_{1,-2} & X_{2,-2} & X_{2,-3} & - X_{1,2} & -X_{2,2} & - X_{3,2} \\
X_{1,-3} & X_{2,-3} & X_{3,-3} & - X_{1,3} & - X_{2,3} & - X_{3,3} 
\end{pmatrix}.
\label{eq:sp6:rep}
\nonumber
\end{equation}
As $\mathfrak{sp}(6, \RR)$ is a simple Lie algebra of rank $3$, it possesses $3$ functionally independent invariants $C_{2}$, $C_{4}$ and $C_{6}$ of degrees $2$, $4$ and $6$, respectively, which can be obtained by the standard determinantal methods \cite{Perelomov1968,Gruber1964}. 
Considering the characteristic equation $|A_{\Gamma_{\omega_{1}}} - \lambda {\rm Id}_6|$ of the matrix, the expansion gives (see \cite{C46})
\begin{equation*}
|A_{\Gamma_{\omega_{1}}} - \lambda {\rm Id}_6|=\lambda^6+ C_2\lambda^4+C_4\lambda^2+ C_6.
\end{equation*}
The quadratic Casimir operator $C_2$, corresponding to the Killing form, has the following symmetric expression  
\begin{equation}
\begin{split}
C_{2} =&  X_{1,1}^{2} + X_{2,2}^{2}  + X_{3,3}^{2} + X_{1,2} X_{2,1}+X_{2,1} X_{1,2}+ X_{1,3}X_{3,1}+X_{3,1} X_{1,3}   \\
& +X_{2,3}X_{3,2}+ X_{3,2}X_{2,3}- X_{-1,2}X_{1,-2}-X_{1,-2}X_{-1,2}  - X_{-1,3}X_{1,-3} \\
& - X_{1,-3}X_{-1,3} -X_{-2,3} X_{2,-3}- X_{2,-3} X_{-2,3}- 2X_{-1,1}X_{1,-1}\\
& -2X_{1,-1}X_{-1,1}-2 X_{-2,2} X_{2,-2}-2X_{2,-2} X_{-2,2} -2 X_{-3,3}X_{3,-3}\\
&-X_{3,-3}X_{-3,3}.
\end{split}
\label{eq:sp6:C2}
\end{equation}
For the purposes of this work, the invariant $C_2$ is sufficient. The explicit expressions for $C_{4}$ and $C_{6}$, that are obtained along the same lines, are omitted due to their length.

The realization $\Phi: \mathfrak{sp}(6, \RR) \to \mathfrak{X}(\T^{*}\RR^{3})$ induced by $\Gamma_{\omega_{1}}$ is spanned by the following $21$ vector fields expressed in global coordinates  $(\mathbf{q}, \mathbf{p})=(q_{1},q_{2}, q_{3},  p_{1},  p_{2},  p_{3})$ of $\T^{*}\RR^{3}$:
 { 
\begin{align}
 & \mathbf{X}_{1} :=\Phi(X_{1,1})  =  q_{1} \pdv{q_{1}} - p_{1} \pdv{p_{1}},& & \mathbf{X}_{2} := \Phi(X_{1,2}) = q_{1} \pdv{q_{2}} - p_{2} \pdv{p_{1}}  \nonumber \\
 & \mathbf{X}_{3}:= \Phi(X_{1,3})= q_{1} \pdv{q_{3}} - p_{3} \pdv{p_{1}}, & & \mathbf{X}_{4} := \Phi(X_{2,1}) = q_{2} \pdv{q_{1}} - p_{1} \pdv{p_{2}} \nonumber \\
& \mathbf{X}_{5} := \Phi (X_{2,2}) = q_{2} \pdv{q_{2}} - p_{2} \pdv{p_{2}},& &  \mathbf{X}_{6}:= \Phi (X_{2,3}) = q_{2} \pdv{q_{3}} - p_{3} \pdv{p_{2}}\nonumber \\
&\mathbf{X}_{7} := \Phi (X_{3,1}) = q_{3} \pdv{q_{1}} - p_{1} \pdv{p_{3}}, & & \mathbf{X}_{8}:= \Phi(X_{3,2}) = q_{3} \pdv{q_{2}} - p_{2} \pdv{p_{3}}\nonumber \\
& \mathbf{X}_{9} := \Phi (X_{3,3}) = q_{3} \pdv{q_{3}} - p_{3} \pdv{p_{3}}, & & \mathbf{X}_{10} := \Phi (X_{-1,1}) = - q_{1} \pdv{p_{1}}\label{eq:realization_sp6} \\
& \mathbf{X}_{11} := \Phi (X_{-1,2})= - q_{2} \pdv{p_{1}} - q_{1} \pdv{p_{2}}, & & \!   \mathbf{X}_{12}:= \Phi(X_{-1,3})=- q_{3} \pdv{p_{1}} - q_{1} \pdv{p_{3}}   \nonumber\\
 & \mathbf{X}_{13} := \Phi (X_{-2,2}) = - q_{2} \pdv{p_{2}}, & &\!  \mathbf{X}_{14} := \Phi(X_{-2,3}) = - q_{3} \pdv{p_{2}} - q_{2} \pdv{p_{3}} \nonumber \\
& \mathbf{X}_{15} := \Phi(X_{-3,3}) = - q_{3} \pdv{p_{3}}, & & \mathbf{X}_{16} := \Phi( X_{1,-1}) = p_{1} \pdv{q_{1}}  \nonumber \\
& \mathbf{X}_{17} := \Phi(X_{1,-2}) = p_{2} \pdv{q_{1}} + p_{1} \pdv{q_{2}},& &\mathbf{X}_{18}:= \Phi(X_{1,-3})= p_{3} \pdv{q_{1}} + p_{1} \pdv{q_{3}} \nonumber \\
&\mathbf{X}_{19} := \Phi(X_{2,-2}) = p_{2} \pdv{q_{2}},  & & \mathbf{X}_{20}:= \Phi(X_{2,-3}) = p_{3} \pdv{q_{2}} + p_{2} \pdv{q_{3}}  \nonumber\\
& \mathbf{X}_{21}:= \Phi(X_{3,-3}) = p_{3} \pdv{q_{3}}. & &\nonumber 
\end{align} }
As these vector fields are naturally associated to the fundamental representation $\Gamma_{\omega_{1}}$ of $\mathfrak{sp}(6, \RR)$, they are Hamiltonian vector fields with respect to the canonical symplectic form 
\begin{equation}
\omega := \rd q_{1} \wedge \rd p_{1} + \rd q_{2} \wedge \rd p_{2} + \rd q_{3} \wedge \rd p_{3}
\label{eq:sp6:symplectic}
\end{equation}
of $\T^{*}\RR^{3}$. The corresponding Hamiltonian functions, determined by means of the inner product condition (\ref{inner}), have the following expressions:

\begin{equation}
\begin{array}{lllll}
& \displaystyle h_{1} = q_{1} p_{1},\qquad  & \displaystyle h_{2} = q_{1} p_{2},\qquad  &\displaystyle h_{3} = q_{1}p_{3},\qquad  & \displaystyle h_{4} = q_{2}p_{1} \\[8pt]
& \displaystyle h_{5} = q_{2}p_{2}, \qquad  &\displaystyle h_{6} = q_{2}p_{3},\qquad  & \displaystyle h_{7} = q_{3}p_{1},\qquad & \displaystyle h_{8} = q_{3}p_{2} \\[8pt]
& \displaystyle h_{9} = q_{3}p_{3},\qquad  & \displaystyle h_{10} = \frac{1}{2} q_{1}^{2},\qquad  &\displaystyle h_{11} = q_{1}q_{2},\qquad  & \displaystyle h_{12} = q_{1}q_{3} \\[8pt]
& \displaystyle h_{13} = \frac{1}{2}q_{2}^{2}, \qquad  &\displaystyle h_{14} = q_{2}q_{3},\qquad  & \displaystyle h_{15} = \frac{1}{2} q_{3}^{2},\qquad & \displaystyle h_{16} = \frac{1}{2}p_{1}^{2} \\[8pt]
& \displaystyle h_{17} = p_{1}p_{2},\qquad  & \displaystyle h_{18} = p_{1}p_{3},\qquad  &\displaystyle h_{19} = \frac{1}{2}p_{2}^{2},\qquad  & \displaystyle h_{20} = p_{2}p_{3} \\[8pt]
& \displaystyle h_{21} = \frac{1}{2}p_{3}^{2}.
\end{array}
\label{eq:hamiltonian_sp6}
\end{equation}
A routine computation shows that  they span a Lie algebra isomorphic to $\mathfrak{sp}(6, \RR)$ with respect to the Poisson bracket $\{\cdot, \cdot\}_{\omega}$ induced on $C^{\infty} (\T^{*} \RR^{3})$ by the canonical symplectic form $\omega$. 

By the procedure described in \cite{Campoamor2024}, the $t$-dependent vector field 
\begin{equation}
\mathbf{X} := \sum_{i = 1}^{21} b_{i}(t) \mathbf{X}_{i}
\label{spX}
\end{equation}
where the $b_{i} \in C^{\infty}(\RR)$ are arbitrary $t$-dependent functions $(1 \leq i \leq 21)$, determines  a LH system on $\T^{*}\RR^{3}$. Its VG Lie algebra $V^{X}$ and LH algebra $\mathcal{H}_{\omega}$, respectively spanned by the vector fields \eqref{eq:realization_sp6}  and the Hamiltonian functions \eqref{eq:hamiltonian_sp6}, are both isomorphic to $\mathfrak{sp}(6, \RR)$. The first-order system of ODEs on $\T^{*}\RR^{3}$ associated to $\mathbf{X}$ is given by 
\begin{equation}
\begin{pmatrix}
\dot{q}_{1} \\[1pt]
\dot{q}_{2} \\[1pt]
\dot{q}_{3} \\[1pt]
\dot{p}_{1} \\[1pt]
\dot{p}_{2} \\[1pt]
\dot{p}_{3}
\end{pmatrix} = 
\begin{pmatrix}
b_{1}(t) & b_{4}(t) & b_{7}(t) &   b_{16}(t) &  b_{17}(t) &  b_{18}(t) \\[1pt]
b_{2}(t) & b_{5}(t) & b_{8}(t) &  b_{17}(t) &  b_{19}(t) &  b_{20}(t) \\[1pt]
b_{3}(t) & b_{6}(t) & b_{9}(t) &   b_{18}(t) &   b_{20}(t) &  b_{21}(t)  \\[1pt]
-b_{10}(t) &- b_{11}(t) &- b_{12 }(t) & -b_{1}(t) & -b_{2}(t) & - b_{3}(t) \\[1pt]
-b_{11}(t) & -b_{13}(t)& - b_{14}(t) & - b_{4}(t)& -b_{5}(t) & - b_{6}(t) \\[1pt]
-b_{12}(t) & -b_{14}(t) &- b_{15}(t) & - b_{7}(t) & - b_{8}(t) & - b_{9}(t) 
\end{pmatrix} \begin{pmatrix}
q_{1} \\[1pt]
q_{2} \\[1pt]
q_{3} \\[1pt]
p_{1} \\[1pt]
p_{2} \\[1pt]
p_{3}
\end{pmatrix}
\label{eq:sp6:system}
\end{equation}
where it is easily seen that the coefficient matrix itself is symplectic. We observe that this Lie system $\mathbf{X}$ cannot be reduced by invariants to a lower-dimensional Lie system, as the maximum rank of the generalized distribution associated to $\mathbf{X}$ is $6$ (see \cite{Campoamor2024} for details on the reduction by invariants of Lie systems).


\subsection{Constants of the motion and superposition rule} \label{subsection:constants}

We now  obtain $t$-independent constants of the motion for the $\mathfrak{sp}(6, \RR)$-LH systems~(\ref{eq:sp6:system})
  by applying the so-called coalgebra formalism~\cite{constants2021,constants2013},  which, in turn,   will    allow us to deduce a (nonlinear) superposition rule.

In terms of the Hamiltonian functions \eqref{eq:hamiltonian_sp6}, spanning the $\mathfrak{sp}(6, \RR)$-LH algebra $\mathcal{H}_{\omega}$, the quadra\-tic Casimir invariant $C_{2}$ in \eqref{eq:sp6:C2} reads
\begin{equation}
\begin{split}
C_{2} =& h_{1}^{2} + h_{5}^{2} + h_{9}^{2} + 2h_{2}h_{4} + 2h_{3}h_{7} + 2h_{6}h_{8} - 2 h_{11} h_{17} - 2 h_{12} h_{18}  \\
& - 2 h_{14} h_{20}- 4 h_{10}h_{16} - 4 h_{13} h_{19} - 4 h_{15} h_{21}.
\end{split}
\label{eq:sp6:CasLH}
\end{equation} 
The smallest integer $s$ such that the diagonal prolongations of the vector fields \eqref{eq:realization_sp6} are linearly independent at a generic point of the product manifold 
\begin{equation}
\bigl(\T^{*}\RR^{3}\bigr)^{s } := \T^{*}\RR^{3} \times \overset{s }{\cdots} \times \T^{*}\RR^{3}
\nonumber
\end{equation}
 is $s = 6$. Thus, we consider the  Hamiltonian functions $h_{i}^{(k)} \in C^{\infty}\bigl((\T^{*}\RR^{3})^{k}\bigr)$, for $1 \leq k \leq 7$ , given by 
\begin{equation}
h_{i}^{(k)} = \sum_{\ell = 1}^{k} h_{i}\bigl(\mathbf{q}^{(\ell)}, \mathbf{p}^{(\ell)}\bigr), \qquad 1 \leq i \leq 21
\nonumber
\end{equation}
where $\bigl(\mathbf{q}^{(\ell)}, \mathbf{p}^{(\ell)}\bigr) $ denote the canonical global coordinates in the $\ell^{\textup{th}}$-copy  of $\T^{*}\RR^{3}$ within the product manifold. Using the Casimir invariant \eqref{eq:sp6:CasLH}, we obtain the constants of the motion $F^{(k)} := C_{2}\bigl(h_{1}^{(k)}, \ldots, h_{21}^{(k)}\bigr)$ for the diagonal prolongation $\widetilde{\mathbf{X}}^{7}$ of $\mathbf{X}$ to $\bigl(\T^{*}\RR^{3}\bigr)^{7}$. Concretely, we find that $F^{(1)} = 0$, while 
\begin{equation}
F^{(2)} = - \left(  p_{1}^{(2)} q_{1}^{(1)} - p_{1}^{(1)} q_{1}^{(2)} + p_{2}^{(2)} q_{2}^{(1)} - p_{2}^{(1)} q_{2}^{(2)} + p_{3}^{(2)} q_{3}^{(1)} - p_{3}^{(1)} q_{3}^{(2)} \right)^{2}.
\nonumber
\end{equation}
This second-order constant of the motion further gives rise   to  the following 10 constants of the motion by permutation $S_{ij}$ of the variables $\bigl(\mathbf{q}^{(i)}, \mathbf{p}^{(i)}\bigr) \leftrightarrow \bigl(\mathbf{q}^{(j)}, \mathbf{p}^{(j)}\bigr)$:
\begin{equation}
\begin{split}
&F_{13}^{(2)} = S_{13}\bigl(F^{(2)}\bigr), \qquad F_{14}^{(2)} = S_{14}\bigl(F^{(2)}\bigr), \qquad F_{15}^{(2)} = S_{15}\bigl(F^{(2)}\bigr)\\ 
& F_{16}^{(2)} = S_{16}\bigl(F^{(2)}\bigr) , \qquad
  F_{17}^{(2)} = S_{17}\bigl(F^{(2)}\bigr), \qquad  F_{23}^{(2)} = S_{23}\bigl(F^{(2)}\bigr) \\
  &F_{24}^{(2)} = S_{24}\bigl(F^{(2)}\bigr), \qquad F_{25}^{(2)} = S_{25}\bigl(F^{(2)}\bigr), \qquad
F_{26}^{(2)} = S_{26}\bigl(F^{(2)}\bigr)\\
&  F_{27}^{(2)} = S_{27}\bigl(F^{(2)}\bigr).
\end{split}
\nonumber
\end{equation}
Note that the third-order constant of the motion $F^{(3)}$ is expressed in terms of the quadratic constants of the motion $F^{(2)}$, $F_{13}^{(2)}$  and $F_{23}^{(2)}$ as 
\begin{equation}
F^{(3)} = F^{(2)} + F_{13}^{(2)} + F_{23}^{(2)}  . 
\nonumber
\end{equation} 
In this situation, a superposition rule for the $\mathfrak{sp}(6, \RR)$-LH system \eqref{eq:sp6:system} in terms of $6$ significative constants $k_\ell$ and $6$ particular solutions $\bigl(\mathbf{q}^{(\ell)}, \mathbf{p}^{(\ell)}\bigr) $, with  $1 \leq \ell \leq 6$, can be derived by solving the system of algebraic equations 
\begin{equation}
\begin{split}
& F^{(2)} = - k_{1}^{2}, \qquad F_{23}^{(2)} = - k_{2}^{2}, \qquad F_{24}^{(2)} = -k_{3}^{2}, \qquad F_{25}^{(2)} = -k_{4}^{2} \\
&F_{26}^{(2)} = - k_{5}^{2}, \qquad F_{27}^{(2)} = - k_{6}^{2}  
\end{split}
\nonumber
\end{equation}
provided that the general solution corresponds to  $ \bigl(\mathbf{q}^{(7)}(t), \mathbf{p}^{(7)}(t)\bigr) $. The explicit expression, even with the use of symbolic computation packages, is computationally cumbersome and quite long, for which reason we omit its detailed derivation.


\subsection{Applications} \label{subsection:applicationssp6}

The generic $t$-dependent Hamiltonian $h$ obtained from the Hamiltonian functions \eqref{eq:hamiltonian_sp6} is characterized by \begin{equation}
h = \sum_{i = 1}^{21} b_{i}(t) h_{i}
\nonumber
\end{equation}
in such a way that the system \eqref{eq:sp6:system} corresponds to the Hamilton equations on $\T^{*}\RR^{3}$ with respect  to the canonical symplectic form \eqref{eq:sp6:symplectic}. In order to find some interesting and manageable physical applications, some restrictions on the generators are required, i.e., considering sets of generators $\mathcal{S}$ of the LH algebra $\mathcal{H}_{\omega} \simeq \mathfrak{sp}(6, \RR)$ such that ${\rm card} (\mathcal{S} ) < 21$. 

Considering for example the subset 
\begin{equation}
\mathcal{S}_{\mathrm{I}} := \{h_{2}, h_{4}, h_{9}, h_{10}, h_{13}, h_{15}, h_{16}, h_{19}, h_{21}\}
\nonumber
\end{equation}
formed by nine generators of the LH algebra $\mathcal{H}_{\omega}$, it follows that the $t$-dependent Hamiltonian $h^{\mathrm{I}}$  is given by 
\begin{equation}
\begin{split}
h^{\mathrm{I}} =& b_{2}(t) q_{1}p_{2} + b_{4}(t) q_{2}p_{1} + b_{9}(t) q_{3}p_{3} + \frac{b_{10}(t)}{2} q_{1}^{2} + \frac{b_{13}(t)}{2}q_{2}^{2} + \frac{b_{15}(t)}{2}q_{3}^{2} \\
& + \frac{b_{16}(t)}{2} p_{1}^{2} + \frac{b_{19}(t)}{2} p_{2}^{2} + \frac{b_{21}(t)}{2}p_{3}^{2}.
\end{split}
\label{eq:sp6:electgen}
\end{equation}
An alternative choice that involves the same number of generators can be taken as 
\begin{equation}
\mathcal{S}_{\mathrm{II}} := \{h_{2}-h_{4}, h_{3}-h_{7}, h_{6} - h_{8}, h_{10}, h_{13}, h_{15}, h_{16}, h_{19}, h_{21}\}
\nonumber
\end{equation}
where the corresponding $t$-dependent Hamiltonian $h^{\mathrm{II}}$ has the form 
\begin{equation}
 \begin{split}
h^{\mathrm{II}} =&  \frac{b_{16}(t)}{2}p_{1}^{2} + \frac{b_{10}(t)}{2} q_{1}^{2} + \frac{b_{19}(t)}{2} p_{2}^{2} + \frac{b_{13}(t)}{2} q_{2}^{2} + \frac{b_{21}(t)}{2} p_{3}^{2} + \frac{b_{15}(t)}{2} q_{3}^{2} \\ 
&+ b_{2}(t) (q_{1}p_{2} - q_{2}p_{1}) + b_{3}(t) (q_{1}p_{3} - q_{3}p_{1}) + b_{6}(t) (q_{2}p_{3} - q_{3}p_{2})
\end{split}
\label{eq:sp6:osc3d}
\end{equation}
provided that $b_4(t)=-b_2(t)$, $b_7(t)=-b_3(t)$ and $b_8(t)=-b_6(t)$. 
In the latter expression, the mechanical meaning of each of the intervening terms is more transparent, suggesting that the Hamiltonian can be used to describe (coupled) systems in either Euclidean or non-Euclidean spaces. 


\subsubsection{A time-dependent electromagnetic field}
Let us now consider $t$-dependent functions $m_{i}(t), e_{i}(t), \gamma(t) \in C^{\infty}(\RR)$, with $1 \leq i \leq 3$, such that 
\begin{equation}
m_{i}(t) > 0\qquad (1 \leq i \leq 3), \qquad \ddot{\gamma}(t) \neq 0, \qquad t \in \RR.
\nonumber
\end{equation}
Define now the $t$-dependent vector field $\mathbf{A}$ on $\RR^{3}$ given by 
\begin{equation}
A_{1} := - \frac{1}{2} q_{2} \gamma(t), \qquad A_{2} := \frac{1}{2} q_{1} \gamma(t), \qquad A_{3} := \frac{1}{2} q_{3} \gamma(t)
\nonumber
\end{equation}
as well as the scalar oscillator potential 
\begin{equation}
\phi := \phi_{1} + \phi_{2} + \phi_{3}, \qquad \phi_{i} := \frac{1}{2} q_{i}^{2} \qquad (1 \leq i \leq 3). 
\nonumber
\end{equation}

Consider the following choices for the $t$-dependent functions $b_{i}(t)$ appearing in the $t$-depen\-dent Hamiltonian \eqref{eq:sp6:electgen}:
\begin{equation}
\begin{array}{llll}
&\displaystyle b_{2}(t) = - \frac{\gamma(t) e_{2}(t)}{2 m_{2}(t)}, \quad  & \displaystyle b_{10}(t) = e_{1}(t) + \frac{\gamma^{2}(t)e_{2}^{2}(t)}{4m_{2}(t)},\quad  &\displaystyle b_{16}(t) = \frac{1}{m_{1}(t)} \\[10pt]
&\displaystyle  b_{4}(t) = \frac{\gamma(t) e_{1}(t)}{2 m_{1}(t)}, \quad &\displaystyle b_{13}(t) = e_{2}(t)+ \frac{\gamma^{2}(t) e_{1}^{2}(t)}{4 m_{1}(t)},\quad  & \displaystyle b_{19}(t) = \frac{1}{m_{2}(t)}  \\[10pt]
&\displaystyle b_{9}(t) = - \frac{\gamma(t) e_{3}(t)}{2m_{3}(t)} ,\quad  & \displaystyle b_{15}(t) = e_{3}(t) + \frac{\gamma^{2}(t) e_{3}^2(t)}{4 m_{3}(t)},\quad  &\displaystyle b_{21}(t) = \frac{1}{m_{3}(t)} .
\end{array}
\nonumber
\end{equation}
The $t$-dependent Hamiltonian $h^{\mathrm{E}}$ obtained from \eqref{eq:sp6:electgen} with these choices turns out to be
\begin{equation}
h^{\mathrm{E}} = \sum_{i = 1}^{3} \left( \frac{1}{2m_{i}(t)} \bigl(p_{i} - e_{i}(t) A_{i} \bigr)^{2} + e_{i}(t) \phi_{i} \right). 
\nonumber
\end{equation}
This describes the motion on $\RR^{3}$ of three particles of time-dependent masses $m_{i}(t)$ and time-dependent electric charges $e_{i}(t)$ $(1 \leq i \leq 3)$ under the action of a time-dependent electromagnetic field. The magnetic field is given by
\begin{equation}
\mathbf{B}= \nabla \times \mathbf{A} = (0, 0, \gamma(t))
\nonumber
\end{equation}
while  the electric field reads
\begin{equation}
\mathbf{E}= - \nabla \phi - \frac{\partial \mathbf{A}}{\partial t} = -\frac{1}{2}\bigr(  2q_{1} - q_{2} \dot{\gamma}(t),   2 q_{2} + q_{1} \dot{\gamma}(t),  2q_{3} + q_{3} \dot{\gamma}(t) \bigl). 
\nonumber
\end{equation}
As we are assuming that $\ddot{\gamma}(t) \neq 0$ for all $t \in \RR$, the electric field $\mathbf{E}$ is also time-dependent. 


\subsubsection{Time-dependent coupled oscillators} \label{subsection:coupledosc}
We next show that the Hamiltonian \eqref{eq:sp6:osc3d} can be considered as a coupled system, using a similar ansatz to that proposed in \cite{Campoamor2024}. To this extent, we first consider the Whitney sum $\T^{*}\RR^{3} = \T^{*}\RR \oplus \T^{*}\RR \oplus \T^{*}\RR$ together with the three canonical projections $\mathrm{pr}_{k}$ $(1 \leq k \leq 3)$ onto its factors: 
\begin{equation}
\mathrm{pr}_{k}: \T^{*}\RR^{3} \to \T^{*}\RR, \qquad (\mathbf{q}, \mathbf{p}) \mapsto (q_{k}, p_{k}), \qquad 1 \leq k \leq 3.
\nonumber
\end{equation}
The $t$-dependent Hamiltonian $h_{i,j}^{\mathrm{1D}}$ on $\T^{*}\RR$ given by 
\begin{equation}
h_{i,j}^{\mathrm{1D}} := \frac{b_{i}(t)}{2} p^{2} + \frac{b_{j}(t)}{2} q^{2}
\nonumber
\end{equation}
determines a LH system on $\T^{*}\RR$ such that the Hamiltonian \eqref{eq:sp6:osc3d} can be rewritten as 
\begin{equation} 
\begin{split}
h =& \mathrm{pr}_{1}^{*}\bigl(h_{16,10}^{\mathrm{1D}} \bigr) + \mathrm{pr}_{2}^{*}\bigl(h_{19,13}^{\mathrm{1D}}\bigr) + \mathrm{pr}_{3}^{*}\bigl(h_{21,15}^{\mathrm{1D}}\bigr)+ b_{2}(t) (q_{1}p_{2} - q_{2}p_{1}) \\  
& + b_{3}(t) (q_{1}p_{3} - q_{3}p_{1}) + b_{6}(t) (q_{2}p_{3} - q_{3}p_{2}).
\end{split}
\nonumber
\end{equation}
This shows that \eqref{eq:sp6:osc3d} describes a  system formed by three 1D $t$-dependent     Hamiltonians (namely, $h_{16,10}^{\mathrm{1D}}$, $h_{19,13}^{\mathrm{1D}}$ and $h_{21,15}^{\mathrm{1D}}$) coupled through the angular momentum terms $q_{1}p_{2}- q_{2}p_{1}$, $q_{1}p_{3} - q_{3}p_{1}$ and $q_{2}p_{3}- q_{3}p_{2}$. 

Several interesting choices can be made for the $t$-dependent functions taking part in the $t$-dependent Hamiltonian \eqref{eq:sp6:osc3d}, yielding   generalized coupled oscillators with different properties. In order to illustrate the situation, let us consider the positive functions $m_{i}(t), k_{i}(t), \gamma_{i}(t) \in C^{\infty}(\RR)$ for $1 \leq i \leq 3$. 

\begin{itemize}
\item[$\bullet$] \textit{A time-dependent Hamiltonian of coupled harmonic oscillators}. Suppose that 
\begin{equation}
k_{i}(t) > \frac{\gamma_{i}^{2}(t)}{4m_{i}(t)}, \qquad t \in \RR, \qquad 1 \leq i \leq 3. 
\nonumber
\end{equation}
Define now 
\begin{equation}
\Omega_{i}(t) := \sqrt{\frac{1}{m_{i}(t)} \left( k_{i} (t) - \frac{\gamma_{i}^{2}(t)}{4m_{i}(t)} \right)}, \qquad t \in \RR, \qquad 1 \leq i \leq 3. 
\nonumber
\end{equation}
Denote by $h^{\mathrm{CHO}}$ the $t$-dependent Hamiltonian obtained from \eqref{eq:sp6:osc3d} after  the following choices:
\begin{equation}
\begin{split}
&b_{16}(t) = \frac{1}{m_{1}(t)}, \qquad b_{10}(t) = m_{1}(t) \Omega_{1}^{2}(t), \qquad b_{19}(t) = \frac{1}{m_{2}(t)} \\
&b_{13}(t) = m_{2}(t) \Omega_{2}^{2}(t), \qquad b_{21}(t) = \frac{1}{m_{3}(t)}, \qquad b_{15}(t) = m_{3}(t) \Omega_{3}^{2}(t).
\end{split}
\nonumber
\end{equation}
Then, $h^{\mathrm{CHO}}$ reads as 
\begin{equation}
\begin{split}
h^{\mathrm{CHO}} = &  \sum_{i = 1}^{3} \left( \frac{1}{2m_{i}(t)}p_{i}^{2} + \frac{1}{2}m_{i}(t) \Omega_{i}^{2}(t)q_{i}^{2} \right) + b_{2}(t) (q_{1}p_{2} - q_{2}p_{1}) \\
& + b_{3}(t) (q_{1}p_{3} - q_{3}p_{1}) + b_{6}(t) (q_{2}p_{3} - q_{3}p_{2}).
\nonumber
\end{split}
\end{equation}
In this case, the three 1D systems $h_{16,10}^{\mathrm{1D}}$, $h_{19,13}^{\mathrm{1D}}$ and $h_{21,15}^{\mathrm{1D}}$ turn out to be  time-dependent harmonic oscillators with time-dependent masses $m_{i}(t)$, time-dependent damping ``constants''  $\gamma_{i}(t)$, time-dependent spring ``constants'' $k_{i}(t)$ and time-dependent frequencies $\Omega_{i}(t)$ $(1 \leq i \leq 3)$. Additional interesting  choices can be considered for the $t$-dependent coupling constants $b_{2}(t)$, $b_{3}(t)$ and $b_{6}(t)$, as they can be taken to depend on the time-dependent masses $m_{i}(t)$,  Hooke ``constants''  $k_{i}(t)$ or   friction ``constants'' $\gamma_{i}(t)$. 

\item[$\bullet$] \textit{Coupled Caldirola--Kanai Hamiltonian}. Let us now define 
\begin{equation}
\lambda_{i}(t) := \frac{\gamma_{i}(t)}{m_{i}(t)}, \qquad \Omega_{i}(t):= \sqrt{\frac{k_{i}(t)}{m_{i}(t)}}, \qquad t \in \RR, \qquad 1 \leq i \leq 3
\nonumber
\end{equation}
and denote by $h^{\mathrm{CCK}}$ the $t$-dependent Hamiltonian obtained from \eqref{eq:sp6:osc3d} after the choices 
\begin{equation}
\begin{split}
&b_{16}(t) = \frac{1}{m_{1}(t)} \re^{-2 \int_{0}^{t} \lambda_{1}(s) \, \rd s}, \qquad b_{10}(t) = m_{1}(t) \Omega_{1}^{2}(t)\re^{2 \int_{0}^{t} \lambda_{1}(s) \,  \rd s} \\
& b_{19}(t) = \frac{1}{m_{2}(t)} \re^{-2 \int_{0}^{t} \lambda_{2}(s)\,  \rd s}, \qquad b_{13}(t) = m_{2}(t) \Omega_{2}^{2}(t)\re^{2 \int_{0}^{t} \lambda_{2}(s) \, \rd s}\\
&b_{21}(t) = \frac{1}{m_{3}(t)} \re^{-2 \int_{0}^{t} \lambda_{3}(s) \,  \rd s}, \qquad b_{15}(t) = m_{3}(t) \Omega_{3}^{2}(t)\re^{2 \int_{0}^{t} \lambda_{3}(s) \,  \rd s}.
\end{split}
\nonumber
\end{equation}
In this case, $h^{\mathrm{CCK}}$ adopts the form 
\begin{equation}
\begin{split}
h^{\mathrm{CCK}} &= \sum_{i = 1}^{3} \left( \frac{1}{2m_{i}(t)}  \re^{-2 \int_{0}^{t} \lambda_{i}(s) \, \rd s} p_{i}^{2} + \frac{1}{2} m_{i}(t) \Omega_{i}^{2}(t)\re^{2 \int_{0}^{t} \lambda_{i}(s) \, \rd s} q_i^2\right)  \\
& + b_{2}(t) (q_{1}p_{2} - q_{2}p_{1}) + b_{3}(t) (q_{1}p_{3} - q_{3}p_{1}) + b_{6}(t) (q_{2}p_{3} - q_{3}p_{2}).
\end{split}
\nonumber
\end{equation}
It is straightforward to verify that the 1D systems $h_{16,10}^{\mathrm{1D}}$, $h_{19,13}^{\mathrm{1D}}$ and $h_{21,15}^{\mathrm{1D}}$ correspond to the so-called Caldirola--Kanai Hamiltonians~\cite{Caldirola1941,Kan}. They describe the motion of a particle of time-dependent mass $m_{i}(t)$  attached to a spring with time-dependent Hooke ``constant'' given by $k_{i}(t) := m_{i}(t) \Omega_{i}^{2}$, with $\Omega_{i}(t)$ being the time-dependent frequency, and subjected to a frictional force with a time-dependent friction ``constant'' $\gamma_{i}(t):= m_{i}(t) \lambda_{i}(t)$ for $1 \leq i \leq 3$.  

\end{itemize}


\section{The special unitary Lie algebra $\mathfrak{su}(3)$} \label{section:su3}
 The relevance of the Lie algebra $\mathfrak{su}(3)$ is widely recognized due to its importance in particle physics (see \cite{Georgi1999} and references therein). From the algebraic point of view, recall that $\mathfrak{su}(3)$ is the compact real form of the (complex) Lie algebra $\mathfrak{sl}(3, \CC)$. Let us now consider the $\mathfrak{su}(3)$-subalgebra of $\mathfrak{sp}(6, \RR)$ spanned by 
\begin{equation}
\begin{split}
&E_{1}^{+} := \frac{1}{2}(X_{1,2} - X_{2,1} - X_{-1,2} + X_{1,-2} )\\
& E_{2}^{+} := \frac{1}{2}(X_{2,3} - X_{3,2} - X_{-2,3} + X_{2,-3}) \\
&E_{1}^{-} := \frac{1}{2}(- X_{1,2} + X_{2,1} - X_{-1,2} + X_{1,-2})\\
& E_{2}^{-} := \frac{1}{2}(-X_{2,3} + X_{3,2} - X_{-2,3} + X_{2,-3}) .
\end{split}
\label{eq:su3:simpleroots}
\end{equation}
The elements $E_{1}^{+}, E_{2}^{+}$ correspond to the two positive simple roots of $\mathfrak{su}(3)$, while $E_{1}^{-}$ and $E_{2}^{-}$ are associated with the negative roots. The generators of the Cartan subalgebra are
\begin{equation}
\begin{split}
&H_{1} := [E_{1}^{+}, E_{1}^{-}] = -X_{-1,1} + X_{-2,2} + X_{1,-1} - X_{2,2} \\
&H_{2} := [E_{2}^{+}, E_{2}^{-}]  = - X_{-2,2} + X_{-3,3} + X_{2,-2} - X_{3,-3}. 
\end{split}
\label{eq:su3:Cartan}
\end{equation}
The  Cartan--Weyl basis of $\mathfrak{su}(3)$ is obtained by from \eqref{eq:su3:simpleroots}  and  \eqref{eq:su3:Cartan} together with the non-simple roots 
\begin{equation}
\begin{split}
&E_{3}^{+} := [E_{1}^{+}, E_{2}^{+}] = \frac{1}{2}(X_{1,3} - X_{3,1} - X_{-1,3} + X_{1,-3}) \\
&E_{3}^{-} :=- [E_{1}^{-}, E_{2}^{-}] = \frac{1}{2}(-X_{1,3} + X_{3,1} - X_{-1,3} + X_{1,-3}).
\end{split}
\nonumber
\end{equation}
The nonvanishing commutation relations of $\mathfrak{su}(3)$ over this basis, up to skew-symmetry, are given by  
\begin{equation}
\begin{array}{llll}
&  [E_{1}^{+}, E_{2}^{+}] = E_{3}^{+},\quad  & [E_{1}^{+}, E_{1}^{-}] = H_{1},\quad  & [E_{1}^{+}, E_{3}^{-}]= - E_{2}^{-}\\[8pt]  
&  [E_{1}^{+}, H_{1}] = - 2E_{1}^{+},  \quad  &[E_{1}^{+}, H_{2}] = E_{1}^{+}, \quad &[E_{2}^{+}, E_{2}^{-}] = H_{2} \\[8pt]
& [E_{2}^{+}, E_{3}^{-}] = E_{1}^{-}, \quad & [E_{2}^{+}, H_{1}] = E_{2}^{+}, \quad & [E_{2}^{+}, H_{2}] = - 2E_{2}^{+} \\[8pt]
&[E_{3}^{+}, E_{1}^{-}]  = - E_{2}^{+}, \quad & [E_{3}^{+}, E_{2}^{-}] = E_{1}^{+}, \quad &[E_{3}^{+}, E_{3}^{-}] = H_{1} + H_{2} \\[8pt]
&[E_{3}^{+}, H_{1}] = - E_{3}^{+}, \quad & [E_{3}^{+}, H_{2}] = - E_{3}^{+}, \quad &[E_{1}^{-}, E_{2}^{-}] = -E_{3}^{-} \\[8pt]
&[E_{1}^{-}, H_{1}] = 2E_{1}^{-}, \quad & [E_{1}^{-}, H_{2}] = - E_{1}^{-}, \quad &[E_{2}^{-}, H_{1}] = - E_{2}^{-} \\[8pt]
&[E_{2}^{-}, H_{2}] = 2E_{2}^{-}, \quad &[E_{3}^{-}, H_{1}] = E_{3}^{-}, \quad &[E_{3}^{-}, H_{2}] = E_{3}^{-}.
\end{array}
\nonumber
\end{equation}
The fundamental representation $\Gamma_{\omega_{1}}$ of $\mathfrak{sp}(6, \RR)$, when restricted to the subalgebra $\mathfrak{su}(3)$, gives rise to the branching rule  $\mathfrak{su}(3)$-representation $\Gamma_{\omega_{1}} \vert_{\mathfrak{su}(3)}: \mathfrak{su}(3) \to \mathfrak{gl}(6, \RR)$ determined by the matrix condition 
\begin{equation}
A_{\Gamma_{\omega_{1}} \vert_{\mathfrak{su}(3)}} = \begin{pmatrix}
1 & 0 \\
0 & 1
\end{pmatrix} \otimes B_{1} + \begin{pmatrix}
0 & 1 \\
1& 0 
\end{pmatrix} \otimes B_{2}
\nonumber
\end{equation}
where $B_{1}$ and $B_{2}$ are the $3\times3$ real matrices  given by
\begin{equation}
\begin{split}
&B_{1} :=  \frac{1}{2} \begin{pmatrix}
0 & E_{1}^{+} - E_{1}^{-} & E_{3}^{+} - E_{3}^{-} \\[2pt]
- E_{1}^{+} + E_{1}^{-} & 0 & E_{2}^{+} - E_{2}^{-} \\[2pt]
- E_{3}^{+} + E_{3}^{-} & - E_{2}^{+} + E_{2}^{-} & 0 
\end{pmatrix}\\[5pt]
&B_{2} := \frac{1}{2} \begin{pmatrix}
2 H_{1} & E_{1}^{+} + E_{1}^{-} & E_{3}^{+} + E_{3}^{-} \\[2pt]
E_{1}^{+} + E_{1}^{-} & - 2H_{1}  + 2H_{2} & E_{2}^{+} + E_{2}^{-} \\[2pt]
E_{3}^{+} + E_{3}^{-}  & E_{2}^{+} + E_{2}^{-} & - 2 H_{2}
\end{pmatrix}.
\end{split}
\nonumber
\end{equation}
In particular, note that the restriction of the fundamental representation $\Gamma_{\omega_{1}}$ of $\mathfrak{sp}(4, \RR)$ to $\mathfrak{su}(3)$ gives rise to the branching rule (see \cite{Patera1981})
\begin{equation}
(1,0,0) \downarrow (1,0) \oplus (0,1)
\nonumber
\end{equation}
where $(1,0,0)$ is highest weight of the representation $\Gamma_{\omega_{1}}$, while $(1, 0)$ and $(0,1)$ are the highest weights of the  three-dimensional quark and antiquark representations of $\mathfrak{su}(3)$, respectively\footnote{The representation $(1, 0) \oplus (0,1)$ is irreducible as a real representation of $\mathfrak{su}(3)$, but reducible as a complex one.}. 
The associated realization $\Psi: \mathfrak{su}(3) \to \mathfrak{X}(\T^{*}\RR^{3})$ induced by the representation $\Gamma_{\omega_{1}} \vert_{\mathfrak{su}(3)}$ is thus spanned by the eight vector fields 
\begin{equation}
\begin{split}
&\mathbf{Y}_{1} :=  \Psi(E_{1}^{+}), \qquad \mathbf{Y}_{2}:=    \Psi(E_{2}^{+}), \qquad \mathbf{Y}_{3} :=    \Psi(E_{3}^{+})\\
&\mathbf{Y}_{4} :=   \Psi(E_{1}^{-}), \qquad \mathbf{Y}_{5}:=    \Psi(E_{2}^{-}), \qquad \mathbf{Y}_{6}:=    \Psi(E_{3}^{-}) \\
&\mathbf{Y}_{7}:=    \Psi(H_{1}), \qquad\, \mathbf{Y}_{8}:=   \Psi(H_{2}).   
\end{split}
\nonumber
\end{equation}
In terms of the global coordinates $(\mathbf{q}, \mathbf{p}) $ of $\T^{*}\RR^{3}$, these vector fields are explicitly given by 
\begin{equation}
\begin{split}
\mathbf{Y}_{1} &= \frac{1}{2} \left( (-q_{2} + p_{2}) \pdv{q_{1}} + (q_{1} + p_{1}) \pdv{q_{2}} + (q_{2} - p_{2}) \pdv{p_{1}} + (q_{1} + p_{1}) \pdv{p_{2}}\right) \\
\mathbf{Y}_{2} & =\frac{1}{2} \left( (-q_{3} + p_{3}) \pdv{q_{2}} + (q_{2} + p_{2}) \pdv{q_{3}} + (q_{3}- p_{3}) \pdv{p_{2}} + (q_{2} + p_{2}) \pdv{p_{3}} \right)\\
\mathbf{Y}_{3} & = \frac{1}{2} \left( (- q_{3} + p_{3}) \pdv{q_{1}} + (q_{1}+ p_{1}) \pdv{q_{3}} + (q_{3} - p_{3}) \pdv{p_{1}} + (q_{1}+ p_{1}) \pdv{p_{3}} \right) \\
\mathbf{Y}_{4} & = \frac{1}{2} \left( ( q_{2}+ p_{2}) \pdv{q_{1}} + (- q_{1}+ p_{1}) \pdv{q_{2}} + (q_{2} + p_{2}) \pdv{p_{1}} + (q_{1}- p_{1}) \pdv{p_{2}} \right) \\
\mathbf{Y}_{5} & = \frac{1}{2} \left( (q_{3} + p_{3}) \pdv{q_{2}} + (- q_{2} + p_{2}) \pdv{q_{3}}+ (q_{3} + p_{3}) \pdv{p_{2}} + (q_{2} - p_{2}) \pdv{p_{3}} \right) \\
\mathbf{Y}_{6} & = \frac{1}{2} \left( (q_{3} + p_{3}) \pdv{q_{1}} + (- q_{1}+ p_{1}) \pdv{q_{3}} + (q_{3} + p_{3}) \pdv{p_{1}} + (q_{1} -p_{1}) \pdv{p_{3}} \right) \\
\mathbf{Y}_{7} & = p_{1} \pdv{q_{1}}  - p_{2} \pdv{q_{2}} + q_{1} \pdv{p_{1}} - q_{2} \pdv{p_{2}}  \\
\mathbf{Y}_{8} & = p_{2} \pdv{q_{2}} - p_{3} \pdv{q_{3}} + q_{2}  \pdv{p_{2}} - q_{3} \pdv{p_{3}}.
\end{split}
\label{eq:su3:vectorfields}
\end{equation}
As expected, they are Hamiltonian vector fields with respect to the canonical symplectic form \eqref{eq:sp6:symplectic}, as they have been obtained by restriction of the fundamental representation $\Gamma_{\omega_{1}}$ of $\mathfrak{sp}(6,\RR)$. The corresponding Hamiltonian functions $h_{i}'$, fulfilling the  inner product condition  (\ref{inner}),      are easily verified to be   
\begin{equation}
\begin{split}
&h_{1}' = \frac{1}{2} (q_{1}p_{2} - q_{2}p_{1} + p_{1}p_{2} - q_{1}q_{2} ), \qquad h_{2}' = \frac{1}{2} (q_{2}p_{3} - q_{3}p_{2} + p_{2}p_{3} - q_{2}q_{3})\\
&h_{3}' = \frac{1}{2}(q_{1}p_{3} - q_{3}p_{1} + p_{1}p_{3} - q_{1}q_{3}), \qquad  h_{4}' = \frac{1}{2}(- q_{1}p_{2} + q_{2}p_{1} + p_{1}p_{2} - q_{1}q_{2}) \\
& h_{5}' = \frac{1}{2}(-q_{2}p_{3} + q_{3}p_{2} + p_{2}p_{3} - q_{2}q_{3}), \quad\ h_{6}' = \frac{1}{2}(-q_{1}p_{3} + q_{3}p_{1} + p_{1}p_{3} - q_{1}q_{3}) \\
&h_{7}' = \frac{1}{2}(- q_{1}^{2} + q_{2}^{2} + p_{1}^{2}  - p_{2}^{2}), \qquad h_{8}' = \frac{1}{2} (- q_{2}^{2} + q_{3}^{2} + p_{2}^{2} - p_{3}^{2}).
\end{split}
\label{eq:su3:ham}
\end{equation}
A routine computation shows that they span a Lie algebra isomorphic to $\mathfrak{su}(3)$. It follows that the $t$-dependent vector field
\begin{equation}
\mathbf{Y} := \sum_{i = 1}^{8} a_{i}(t) \mathbf{Y}_{i}
\label{eq:su3:tvf}
\end{equation}
where $a_{i} \in C^{\infty}(\RR)$ is an arbitrary $t$-dependent function $(1 \leq i \leq 8)$, determines a LH system on $\T^{*}\RR^{3}$. Indeed, its  VG Lie algebra $V^{Y}$, spanned by the vector fields \eqref{eq:su3:vectorfields}, and  its LH algebra $\mathcal{H}_{\omega}'$, spanned by the Hamiltonian functions \eqref{eq:su3:ham}, are both isomorphic to $\mathfrak{su}(3)$. 

It is worthy to be observed that the $\mathfrak{su}(3)$-Lie system $\mathbf{Y}$ cannot be reduced by invariants to a lower-dimensional system, as the subalgebra $\mathfrak{su}(3) \subset \mathfrak{sp}(6, \RR)$ is irreducibly embedded and the $\mathfrak{sp}(6, \RR)$-Lie system $\mathbf{X}$ (\ref{spX}) cannot itself be reduced by invariants, as pointed out before (see \cite{Campoamor2018} for more details). It should be observed that a superposition rule for the $\mathfrak{su}(3)$-LH system \eqref{eq:su3:tvf} can be derived following the same procedure described in Subsection~\ref{subsection:constants} for the $\mathfrak{sp}(6, \RR)$-LH system \eqref{eq:sp6:system}.


\subsection{Applications to  coupled systems} \label{subsection:appsu3}
Following an approach similar to that of Subsection~\ref{subsection:applicationssp6}, we first look for a minimal set of generators of the $\mathfrak{su}(3)$-LH algebra $\mathcal{H}_{\omega}'$. A simple inspection shows that $h_{1}'$, $h_{2}'$, $h_{4}'$ and $h_{5}'$ span $\mathcal{H}_{\omega}'$, as they are associated to the two simple roots of the subalgebra $\mathfrak{su}(3) \subset \mathfrak{sp}(6, \RR)$. Among all possible linear combinations of these generators, let us consider the following new set of generators of $\mathcal{H}_{\omega}'$: 
\begin{equation}
\begin{split}
&\tilde{h}_{1} := h_{1}' + h_{4}' = p_{1}p_{2} - q_{1}q_{2}, \qquad \tilde{h}_{2} := h_{1}' - h_{4}' = q_{1}p_{2} - q_{2}p_{1} \\
&\tilde{h}_{3} := h_{2}' + h_{5}' = p_{2}p_{3} - q_{2}q_{3}, \qquad \tilde{h}_{4} := h_{2}' - h_{5}' = q_{2}p_{3} - q_{3}p_{2}. 
\end{split}
\nonumber
\end{equation}
The $t$-dependent Hamiltonian $\tilde{h}$ corresponding to this choice of generators reads  
\begin{equation}
\begin{split}
\tilde{h} =& \tilde{a}_{1}(t) (p_{1}p_{2} - q_{1}q_{2}) + \tilde{a}_{2}(t) (q_{1}p_{2} -q_{2}p_{1})  \\
& + \tilde{a}_{3}(t) (p_{2}p_{3} - q_{2}q_{3}) + \tilde{a}_{4}(t) (q_{2}p_{3} -q_{3}p_{2}). 
\end{split}
\label{hamsu3}
\end{equation}
Alternatively, $\tilde{h}$ can also be obtained from the $t$-dependent Hamiltonian $h' := \sum_{i = 1}^{8} a_{i}(t) h_{i}'$ with the following identifications of the $t$-dependent functions $a_{i}$: 
\begin{equation}
\begin{split}
&a_{1} = \tilde{a}_{1} + \tilde{a}_{2}, \qquad a_{2} = \tilde{a}_{3} + \tilde{a}_{4}, \qquad a_{4} = \tilde{a}_{1} - \tilde{a}_{2}, \qquad a_{5} = \tilde{a}_{3} - \tilde{a}_{4} \\
&a_{3} = a_{6} = a_{7} = a_{8} = 0. 
\end{split}
\nonumber
\end{equation}
The  Hamilton equations associated to $\tilde{h}$ allow the following matrix form 
\begin{equation}
\begin{pmatrix}
\dot{q}_{1} \\[1pt]
\dot{q}_{2} \\[1pt]
\dot{q}_{3} \\[1pt]
\dot{p}_{1} \\[1pt]
\dot{p}_{2} \\[1pt]
\dot{p}_{3}
\end{pmatrix} = 
\begin{pmatrix}
0 & - \tilde{a}_{2}(t) & 0 & 0 & \tilde{a}_{1}(t) & 0 \\[1pt]
\tilde{a}_{2}(t) & 0 & - \tilde{a}_{4}(t) & \tilde{a}_{1}(t) & 0 & \tilde{a}_{3}(t) \\[1pt]
0 & \tilde{a}_{4}(t) & 0 & 0 & \tilde{a}_{3}(t) & 0 \\[1pt]
0 & \tilde{a}_{1}(t) & 0 & 0 & - \tilde{a}_{2}(t) & 0 \\[1pt]
\tilde{a}_{1}(t) & 0 & \tilde{a}_{3}(t) & \tilde{a}_{2}(t) & 0 & - \tilde{a}_{4}(t) \\[1pt]
0 & \tilde{a}_{3}(t) & 0 & 0 &  \tilde{a}_{4}(t) & 0 
\end{pmatrix} \begin{pmatrix}
q_{1} \\[1pt]
q_{2} \\[1pt]
q_{3} \\[1pt]
p_{1} \\[1pt]
p_{2} \\[1pt]
p_{3}
\end{pmatrix}.
\label{eq:su3:system}
\end{equation}
A remarkable fact is that $\tilde{h}$ can be interpreted as a non-trivially  coupled system formed by two subsystems defined on the Minkowskian plane $\mathbf{M}^{1+1}$. Recall that $\mathbf{M}^{1+1}$ is the real plane equipped with the Lorentzian metric $\rd s^{2} = \rd q_{x}^{2} - \rd q_{y}^{2}$, where $(q_{x}, q_{y})$ are global coordinates on $\mathbf{M}^{1+1}$. Using the standard light-cone coordinates 
\begin{equation}
q_{+} = q_{x} + q_{y}, \qquad q_{-} = q_{x} - q_{y}
\nonumber 
\end{equation}
the Minkowskian metric reads as $\rd s^{2} = \rd q_{+} \rd q_{-}$. The cotangent bundle $\T^{*}\mathbf{M}^{1+1}$ is naturally equipped with the canonical global coordinates $(q_{\pm}, p_{\pm})$, where $p_{\pm}$ are the conjugate momentum associated to $q_{\pm}$, that is,
 \begin{equation}
p_{+} = \frac 12 (p_{x} + p_{y}), \qquad p_{-} = \frac 12 (p_{x} - p_{y}).
\nonumber 
\end{equation}
  It follows that the kinetic energy associated to the Minkowskian metric in the light-cone coordinates $(q_{\pm}, p_{\pm})$ is $2p_{+}p_{-}$, while $q_{+}q_{-}$ is just the potential of the isotropic oscillator on $\mathbf{M}^{1+1}$, which is one of the widely studied Drach potentials \cite{Perelomov1990,Ranada2001}. 
As in Subsection~\ref{subsection:coupledosc}, we now describe some coupled LH systems associated to the compact Lie algebra $\mathfrak{su}(3)$. Let us first  consider the projections
\begin{equation}
\begin{split}
&\mathrm{pr}_{12}: \T^{*}\RR^{3} \to \T^{*}\mathbf{M}^{1+1}, \qquad (\mathbf{q}, \mathbf{p}) \mapsto (q_{+} = q_{1}, q_{-} = q_{2}, p_{+} = p_{1}, p_{-} = p_{2}) \\
&\mathrm{pr}_{23}: \T^{*}\RR^{3} \to \T^{*}\mathbf{M}^{1+1}, \qquad (\mathbf{q}, \mathbf{p})  \mapsto (q_{+} = q_{3}, q_{-} = q_{2}, p_{+} = p_{3}, p_{-} = p_{2}).
\end{split}
\nonumber
\end{equation}
Then, the $t$-dependent Hamiltonian $h_{ij}^{\mathrm{2D}}$ on $\T^{*}\mathbf{M}^{1+1}$ given by 
\begin{equation}
h_{ij}^{\mathrm{2D}} := \tilde{a}_{i}(t) (p_{+}p_{-} - q_{+}q_{-}) + \tilde{a}_{j}(t) (q_{+}p_{-} - q_{-}p_{+})
\label{eq:su3:2Dlight}
\end{equation}
is a LH system on $\T^{*}\mathbf{M}^{1+1}$ with respect to the canonical symplectic form $\omega=   \rd q_{+} \wedge \rd p_{+} + \rd q_{-} \wedge \rd p_{-}$ and such that the Hamiltonian $\tilde{h}$  \eqref{hamsu3}  associated to the $\mathfrak{su}(3)$-LH system  \eqref{eq:su3:system}   can be expressed as 
\begin{equation}
\tilde{h} = \mathrm{pr}_{12}^{*}\bigl(h_{12}^{\mathrm{2D}}\bigr) +  \mathrm{pr}_{23}^{*}\bigl(h_{34}^{\mathrm{2D}}\bigr)
\nonumber
\end{equation}
thus corresponding to a coupling of two LH systems defined on $\T^{*}\mathbf{M}^{1+1}$. 

The Hamiltonian $h^{\mathrm{2D}}_{ij}$ (\ref{eq:su3:2Dlight})   is expressed in the usual canonical coordinates $(q_{x}, q_{y}, p_{x}, p_{y})$ of $\T^{*}\mathbf{M}^{1+1}$ as 
\begin{equation}
h_{ij}^{\mathrm{2D}} = \frac{\tilde{a}_{i}(t)}{4} \bigl(p_{x}^{2} - p_{y}^{2}\bigr) -  {\tilde{a}_{i}(t)} \bigl(q_{x}^{2} - q_{y}^{2}\bigr) - \tilde{a}_{j}(t) (q_{x} p_{y} - q_{y}p_{x}).
\nonumber
\end{equation}
This shows that $h_{ij}^{\mathrm{2D}}$ is a particular case of the $\mathfrak{so}(1,3)$-LH systems on $\T^{*}\mathbf{M}^{1+1}$ studied in \cite{Campoamor2024}, where it was shown that $h_{ij}^{\mathrm{2D}}$ actually corresponds to a coupling of two oscillator LH systems defined on $\T^{*}\RR$ through the angular momentum term $q_{x}p_{y}- q_{y}p_{x}$. More precisely, $h_{ij}^{\mathrm{2D}}$ (\ref{eq:su3:2Dlight}) can be written as
\begin{equation}
h_{ij}^{\mathrm{2D}} = \mathrm{pr}_{1}^{*}\bigl(h_{i}^{\mathrm{1D}}\bigr) - \mathrm{pr}_{2}^{*}\bigl(h_{i}^{\mathrm{1D}}\bigr) - \tilde{a}_{j}(t) (q_{x}p_{y}- q_{y}p_{x})
\label{eq:su3:2D}
\end{equation}
where $\mathrm{pr}_{1}: \T^{*}\mathbf{M}^{1+1} \to \T^{*}\RR$ and $\mathrm{pr}_{2}: \T^{*}\mathbf{M}^{1+1} \to \T^{*}\RR$ are the projections of the Whitney sum $\T^{*}\mathbf{M}^{1+1} = \T^{*}\RR \oplus \T^{*}\RR$ onto the first and second factors, respectively, while $h_{i}^{\mathrm{1D}}$ is the $t$-dependent Hamiltonian on $\T^{*}\RR$ given by 
\begin{equation}
h_{i}^{\mathrm{1D}} = \frac{\tilde{a}_{i}(t)}{4}\bigl ( p^{2}-4q^{2} \bigr). 
\label{eq:su3:1D}
\end{equation}
Thus, combining the decompositions \eqref{eq:su3:ham} and \eqref{eq:su3:2D}, we conclude that the $\mathfrak{su}(3)$-LH system \eqref{eq:su3:system} on $\T^{*}\RR^{3}$ can be interpreted as a coupling of two LH systems on $\T^{*}\mathbf{M}^{1+1}$ given by the  Hamiltonians $h_{12}^{\mathrm{2D}}$ and $h_{34}^{\mathrm{2D}}$ in \eqref{eq:su3:2Dlight}, with each one of these corresponding to a coupling of two LH systems on $\T^{*}\RR$ with $t$-dependent Hamiltonians $h_{1}^{\mathrm{1D}}$, $h_{2}^{\mathrm{1D}}$ and $h_{3}^{\mathrm{1D}}$, $h_{4}^{\mathrm{1D}}$  obtained from \eqref{eq:su3:1D}.


\section{Final remarks}

It has been shown that the formal construction of LH systems based on representation theoretical grounds and with underlying VG algebra isomorphic to $\mathfrak{sp}(4,\mathbb{R})$ developed in \cite{Campoamor2024} can be naturally extended to higher rank. In particular, several of the two-dimensional LH systems obtained in that reference can be obtained by restriction of the three-dimensional systems associated to  $\mathfrak{sp}(6,\mathbb{R})$, as follows from the canonical embedding chain of symplectic Lie algebras. In particular, the three-dimensional version of time-dependent electromagnetic fields and coupled oscillators, as coupled harmonic oscillators and coupled Caldirola--Kanai systems~\cite{Caldirola1941,Kan}, have been obtained. On the other hand, restricted LH systems related to the subalgebra embedding $\mathfrak{su}(3)\subset \mathfrak{sp}(6,\mathbb{R})$ have been considered, illustrating how the branching rules of (semisimple) Lie algebras can be used to determine new systems associated to distinctive subalgebras. Formally, there is no obstruction to generalize these results to the generic symplectic algebra $\mathfrak{sp}(2n,\mathbb{R})$ for $n>3$, as well as to relevant semisimple subalgebras of the latter, although the explicit derivation of the corresponding superposition rules is expected to be computationally demanding. Another question to be analyzed in the future is to find suitable conditions on the $t$-dependent parameters that allow us to determine an explicit solution of the LH system, specially in connection with physically relevant cases, such as the coupled oscillators or systems with electromagnetic fields. Work in these directions are currently in progress.


\section*{Acknowledgements}

 \addcontentsline{toc}{section}{Acknowledgements}

This work has been supported by Agencia Estatal de Investigaci\'on (Spain)   under   grant PID2023-148373NB-I00 funded by MCIN /AEI /10.13039/501100011033 / FEDER, UE. O.C. acknowledges a fellowship (grant C15/23) supported by Universidad Complutense de Madrid and Banco de Santander. F.J.H.~acknowledges support  by the  Q-CAYLE Project  funded by the Regional Government of Castilla y Le\'on (Junta de Castilla y Le\'on, Spain) and by the Spanish Ministry of Science and Innovation (MCIN) through the European Union funds NextGenerationEU (PRTR C17.I1).  The authors also acknowledge the contribution of RED2022-134301-T funded by MCIN/AEI/10.13039/501100011033 (Spain).



 \end{document}